\documentclass[preprint2]{aastex}

\usepackage{epsfig,graphicx}
\usepackage{natbib}
\bibpunct{(}{)}{;}{a}{}{,}    

\begin{document}

\title{PHYSICS OF POLARIZED SCATTERING\\ AT MULTI-LEVEL ATOMIC SYSTEMS}
\shorttitle{PHYSICS OF POLARIZED SCATTERING}

\author{ J. O. Stenflo$^{1,2}$} 
\affil{$^1$Institute of Astronomy, ETH Zurich, CH-8093 Zurich, Switzerland }
\affil{$^2$Istituto Ricerche Solari Locarno, Via Patocchi, 6605 Locarno-Monti, Switzerland}

\email{stenflo@astro.phys.ethz.ch} 

\begin{abstract}
The symmetric peak observed in linear polarization in the core of the
solar sodium D$_1$ line at 5896\,\AA\ has remained enigmatic since its
discovery nearly two decades ago. One reason is that the theory of
polarized scattering has not been experimentally tested for
multi-level atomic systems in the relevant parameter domains, although
the theory is continually being used for the interpretation of
astrophysical observations. A laboratory experiment that was set up a decade ago to find
out whether the D$_1$ enigma is a problem of solar physics or quantum
physics revealed that the D$_1$ system has a rich polarization
structure in situations where standard scattering theory predicts
zero polarization, even when optical pumping of the $m$ state
populations of the hyperfine-split ground state is accounted for. Here
we show that the laboratory 
results can be modeled in great quantitative detail if
the theory is extended to include the coherences in
both the initial and final states of the scattering process. Radiative
couplings between the allowed dipole transitions generate coherences
in the initial state. Corresponding coherences in the
final state are then demanded by a phase 
closure selection rule. The experimental results for the well understood
D$_2$ line are used to constrain the two free parameters of the
experiment, collision rate and 
optical depth, to suppress the need for free parameters when fitting
the D$_1$ results. 
\end{abstract}


\maketitle


\section{Introduction}\label{sec:intro}
The development of new theories will often go astray unless guided
and tested by observations or experiments. In the early stages of
quantum physics experiments with polarized scattering helped guide the
development of the theory. The experiments by Wilhelm Hanle in
G\"ottingen \citep{stenflo-hanle24} led not only to the discovery of the well used Hanle
effect but at the same time demonstrated the concept of coherent
superposition of quantum states and the partial decoherence caused by
external magnetic fields. The new quantum theory was built on these
concepts. 

The literature on experiments with the scattering of polarized light
came to an abrupt end around 1935, apparently because the topic was considered
exhausted and less rewarding than other areas, and because the observed
phenomena were believed to be sufficiently understood. The theoretical understanding
covered the parameter domain accessible at that time, which in terms
of polarimetric sensitivity was unbelievably crude in comparison with
what is now possible with modern technology. Therefore the quantum
scattering theory never got experimentally tested over a wide range of
weak polarization phenomena that are accessible with current
instrumentation. The theory is nevertheless being
systematically applied for the interpretation of astrophysical
observations in this untested parameter domain. 

With the implementation in the 1990s of a new technology (ZIMPOL) in solar
spectro-polarimetry
\citep{stenflo-povel95,stenflo-povel01,stenflo-gandetal04}, which
allowed the recording of the Sun's 
spectrum with a polarimetric precision of $10^{-5}$, the linearly
polarized spectrum (Stokes $Q/I$) observed near the solar limb was found to
be as spectrally structured as the intensity spectrum, but the
spectral structures were totally different, it was like uncovering a
new spectral face of the Sun \citep{stenflo-sk96,stenflo-sk97}. It is
therefore referred to 
with the term `Second Solar Spectrum', or with the acronym SS2
(implying that the ordinary intensity spectrum is SS1). The origin of this polarized
spectrum is coherent scattering processes in the
Sun's atmosphere. Magnetic fields do not produce this spectrum, but
they modify it --- SS2 is a playground for the Hanle
effect. 

Through intense theoretical efforts most of these unfamiliar
structures could be identified and understood within the framework of standard
quantum scattering theory, revealing remarkable polarization
signatures of quantum interference between states of different total
angula momenta ($J$ quantum numbers), hyperfine structure and isotope
effects, optical pumping, molecular scattering, and partial frequency redistribution
effects in polarized radiative transfer \citep[cf.][for a review]{stenflo-s04}. 

A culmination of the theoretical work to build a comprehensive
quantum-mechanical foundation for the radiation-matter interactions
occurred with the publication of the monumental monograph by
\citet{stenflo-lanlan04}. A 
regular series of International Workshops has been devoted to the
subject of `Solar Polarization' since 1995, the latest, no.~7, in
Kunming, China, September 2013 \citep{stenflo-spw7book}. The physics
of polarized scattering has been a central topic of these Workshops. 

In spite of these various efforts, one kind of observed feature in the
Second Solar Spectrum has remained enigmatic and stubbornly resisted
all attempts to explain it, namely the weak linear polarization peak
observed at the center of the Na\,{\sc i} D$_1$ line at 5896\,\AA\ and
the Ba\,{\sc ii} D$_1$ line at 4934\,\AA\
\citep{stenflo-setal00a,stenflo-setal00b}, although  attention to the
enigmatic sodium D$_1$ feature was drawn already in the 1996 paper in
{\it Nature} \citep{stenflo-sk96}. The problem is that D$_1$ type lines, which have
angular momentum quantum numbers $J=1/2$ for both the lower and
upper atomic level is expected to be intrinsically unpolarizable,
representing polarization `null' lines. However, since both sodium and barium have
nuclear spin 3/2, both the lower and upper levels are split into
hyperfine structure states with total angular momentum quantum numbers
$F=1$ and 2. A number of attempts have been made to explain the
observed D$_1$ polarization in terms of optical pumping of the
hyperfine structure levels of the ground state
\citep{stenflo-landi98,stenflo-casinietal02,stenflo-casinimanso05},
but all of them 
have failed because the predicted D$_1$ polarization is too small by
about two orders of magnitude and has the wrong symmetry
(anti-symmetric polarization profile instead of the observed symmetric
one)
\citep{stenflo-trujetal02,stenflo-kerkenibommier02,stenflo-klementstenflo03}. 

To find out if this enigma is a problem of solar physics or of quantum
physics, a laboratory experiment was set up to explore the
polarization structure of D$_1$ scattering under controlled
conditions. The experiment unambiguously showed that there is indeed a
problem with the quantum theory of polarized scattering, since the
theory predicts null results when the experiment revealed a rich
polarization structure \citep{stenflo-thalmannspw4,stenflo-thalmannspw5}. 

In search for ways in which the theory of quantum scattering would
need to be extended to explain what was seen in the laboratory
experiment, it was realized that if coherences in {\it both} the
initial {\it and} final ground states could be included in the theory
for the scattering process, then the atomic system would possess a
much richer resonant structure that would be able to generate
polarization signatures of the observed kind \citep{stenflo-stenflospw5}. 

In the present paper we develop these ideas into a phenomenological
extension of current scattering theory and use it to successfully model the observed D$_1$
polarization in great quantitative detail. Through use of experimental
data in the well-understood D$_2$ line to constrain the values of the
collisional and optical depth parameters $\gamma_c$ and $\tau$ it
becomes possible to avoid the use of adjustable parameters when fitting the D$_1$ data.


\section{Laboratory experiment}\label{sec:labexp}
After initial attempts failed to do the laboratory exploration of D$_1$
scattering physics for the sodium D$_1$ line with the use of a
broad-band light source, we realized that a tunable laser was required
to achieve the needed S/N ratio with clean separation between the
D$_2$ and D$_1$ transitions and insignificant stray light. In addition
the tuning gives us the precise shapes of the polarization profiles. 

We then
chose to do the experiment for potassium instead, since solid-state
tunable lasers are not available for the sodium wavelength range. The
K\,{\sc i} D$_2$ 7665\,\AA\ and the K\,{\sc i} D$_1$ 7699\,\AA\ lines
have the same quantum number structure as the Na\,{\sc i} D$_2$
5890\,\AA\ and the Na\,{\sc i} D$_1$ 5896\,\AA\ lines, including
nuclear spin and hyperfine structure, so it is the same physics that
is being tested with potassium as for sodium. 

The set-up has been described in detail by 
\citet{stenflo-thalmannspw4,stenflo-thalmannspw5} together with
examples of the experimental results. Here we 
provide a summary of what is relevant for the understanding of the theoretical
modeling of the data. The heart of
the system is a cross-shaped glass cell that contains the potassium
gas. The cell was tailor-made for our experiment by the late
Alessandro Cacciani in Rome, who had perfected the art of building
such cells for use in magneto-optical narrow-band filters used in
helioseismology and solar magnetometry
\citep{stenflo-cacc78,stenflo-cacc97}. Metallic potassium in 
the stem of the cell is heated to give potassium gas at a temperature
of 100$^\circ$ C at the cell center, where the scattering takes
place. An argon buffer gas is used to suppress diffusion of the
potassium vapor towards the cooler entrance and exit windows, to avoid
deposits that could make the windows opaque. 

The spectral band width of the laser beam is not exactly known,
  but it is at least less than 2\,m\AA. This is much smaller than the
  separation between the $F=1$ and $F=2$ hyperfine levels of the
  ground state, which is 9.1\,m\AA, but it is comparable to the
  corresponding hyperfine splitting of the upper state of the K D$_1$
  line, which is 1.14\,m\AA.

The expanded laser beam is passed through polarizers that allow us to
select one of six states of 100\,\%\ fractional polarizations in
Stokes $Q$, $U$, or $V$ before the beam enters the cell: 100\,\%\ $\pm Q/I$,
$\pm U/I$, or $\pm V/I$. Positive Stokes $Q$ is defined to be linear
polarization oriented perpendicular to the scattering plane. The scanning can be
done with either of two separate laser heads, one for tuning around
the K D$_2$ resonance, the other for tuning around the K D$_1$
resonance. 

In the output arm that receives light scattered by the cell at
90$^\circ$, a piezoelastic modulator followed
by a linear polarizer converts the polarization information (Stokes
$Q$, $U$, or $V$) into intensity modulation, at 84\,kHz for $Q$ and
$U$, at 42\,kHz for $V$, while Stokes $I$ is represented by the
unmodulated (DC) signal. A photomultiplier feeds the signal to a
lock-in amplifier, which demodulates the AC component. The combination
of 6 alternative input states with three alternative output states for
the polarization gives us 18 possible combinations. For each of them
the laser tuning gives us the precise profile shape with m\AA\
resolution. 

Furthermore a set of Helmholtz coils mounted on the cell allows us to
impose on the cell center an external magnetic field with any strength
in the range between $\pm 30$\,G, with an orientation that can be
chosen to be in either of the three spatial directions: perpendicular
to the scattering plane, parallel to the input beam, or parallel to
the scattered beam. 

Because the available parameter space of the experiment is so large,
time and manpower only allowed the experiment to be carried out for some
strategically selected combinations among the various
possibilities. In the present paper we focus on the interpretation of
the transverse field case (field perpendicular to the scattering
plane) for the two combinations (1) Input 100\,\%\ $Q/I$ polarization,
output detection of Stokes $Q$, and (2) input 100\,\%\ $V/I$ polarization,
output detection of Stokes $V$. This selection was done not only
because these cases were most 
thoroughly measured, but because conventional scattering theory
predicts null results for case (1) and nearly null for case (2). This
leads us to identify the observed polarization effects for these
cases as due to previously overlooked, neglected physics. 


\section{Physics of polarized scattering}\label{sec:phys}

\subsection{Theoretical overview}\label{sec:theory}
Let us consider a scattering transition between the magnetic substates
labeled $a$ (initial state), $b$ (intermediate), and $f$ (final) and
introduce the area-normalized complex profile function for radiative
absorption: 
\begin{equation}
\Phi_{ba}={2/i\over\omega_{ba}-\omega -i\gamma/2}
\,,\label{eq:phimmu}
\end{equation}
where 
\begin{equation}
\omega_{ba}=(E_b-E_{a})/\hbar
\label{eq:ommmu}\end{equation}
represents the transition frequency between the energy levels of the
upper and lower magnetic substates, while $\omega$ is the frequency of
the incident radiation. Then the Kramers-Heisenberg formula for the
probability amplitude for the scattering process $a\to
b\to f$ may be written as 
\begin{equation}
w_{\alpha\beta}\sim \langle\,f\,\vert\,{\hat{\bf r}}\cdot
{\bf e}_\alpha\,\vert\,b\,\rangle\,\langle\,b\,\vert\,{\hat{\bf r}}\cdot
{\bf e}_\beta\,\vert\,a\,\rangle\,\Phi_{m\mu_a}\label{eq:scatkh}
\end{equation}
\citep[cf.][]{stenflo-book94,stenflo-s98}. 

The matrix transition elements contain the scalar products between the
dipole moment operator (proportional to the position vector ${\bf r}$)
and the unit vectors ${\bf e}_\beta$ and ${\bf e}_\alpha$ of the
incident and scattered radiation field, respectively. Vector ${\bf r}$
is decomposed in the three spherical vector components labeled by
$q=0,\pm 1$,
which equal the difference $\Delta m$ between the magnetic quantum
numbers of the lower and upper states (with the quantization axis
defined to be 
along the magnetic field vector), while the polarization components of
the radiation field are real linear unit vectors that reside in the
plane that is perpendicular to the propagation direction. 

For classical scattering the probability amplitude is represented by the Jones
matrix 
\begin{equation}
w_{\alpha\beta}\sim\Phi_{-q}\,\varepsilon^{\alpha\ast}_q\,\varepsilon^\beta_q\,,\label{eq:scatclass}
\end{equation}
where the $\varepsilon$ factors embody the scalar products between
${\bf r}$ and ${\bf e}_{\alpha,\,\beta}$
\citep[cf.][]{stenflo-book94}. With our definition of $q$, we
  have  $\Phi_{-q}=\Phi_{ba}$ for the case of a $J=0\to 1$ absorption
  transition. In the general
quantum-mechanical case we can write the Kramers-Heisenberg amplitude
in the following analogous form, to bring out the similarlities and
differences with respect to the classical case as transparently as possible: 
\begin{equation}
w_{\alpha\beta}\sim\Phi_{ba}\,\varepsilon^{\alpha\ast}_{q^\prime}\,\varepsilon^\beta_q\,t_{fb}\,t_{ba}\,.\label{eq:scatgen}
\end{equation}
Here $t_{ba}$ and $t_{fb}$ represent the transition amplitudes for
absorption and emission between the respective magnetic
substates. They are all real quantities, but their relative signs are
of importance. All the other factors in Eq.~(\ref{eq:scatgen}) are in
general complex-valued quantities. 

Since the magnetic quantum numbers for the $a$ and $f$
states may differ, the corresponding $q$ indices may also differ,
which is the reason for using a prime for $q$ in the case of the
emission transition. In the classical case $q^\prime =q$, and the
three resonances (corresponding to the three $q$) have equal
probability amplitude and are driven in phase by the incident
radiation field. In the quantum case the scattering amplitudes get
weighted by the relative complex amplitudes $c_a$ of the initial state
$a$. 

To go from the scattering probability amplitudes $w_{\alpha\beta}$,
which represent the elements of a complex $2\times 2$ matrix $\bf w$,
to obtain the scattering probabilities for the different polarization
states, one forms the bilinear products
$w_{\alpha\beta}\,w_{\alpha^\prime\beta^\prime}^\ast$, which are
elements of a tensor product of type ${\bf w}\otimes {\bf w}^\ast$
with appropriate summations over the various atomic states. 

The standard approach in quantum scattering theory has been the `sum
over histories': To obtain the total probability amplitude for
scattering from state $a$ to state $f$, one sums over the probability
amplitudes for all the possible intermediate states, while keeping the
initial and final states $a$ and $f$ fixed. One then forms the
bilinear product to obtain the probabilities, and afterwards sums over
all the possible initial and final states $a$ and $f$. This results in a
complex $4\times 4$ coherency matrix {\bf W} of the form 
\begin{equation}
{\bf W}\,=\,\sum_a \vert\, c_a\vert^2\,\sum_f\,\Bigl(\,\sum_b {\bf
  w}\,\Bigr)\otimes\Bigl(\,\sum_{b^\prime} {\bf w}^\ast\,\Bigr)\label{eq:wmatold}
\end{equation}
\citep{stenflo-book94,stenflo-s98}. 

There is a very fundamental distinction between the various summations
here: While the sum over the intermediate states is {\it coherent}
(done as a linear superposition of the probability amplitudes), the
sums over the initial and final states are {\it incoherent} (done over
the bilinear products that represent probabilities), weighted by the
relative populations $\vert\, c_a\vert^2$ of the initial states. The
quantum interference phenomena are then exclusively due to the cross
products that arise when the sums over $b$ and $b^\prime$ are multiplied with
each other. 

In the present paper we will present experimental and theoretical
evidence that demonstrates that 
Eq.~(\ref{eq:wmatold}) with its separation between coherent and
incoherent summations is incorrect, and that instead all summations
should be done coherently 
\citep[as previously suggested in][]{stenflo-stenflospw5}. The correct
expression should therefore be 
\begin{equation}
{\bf W}\,=\,\Bigl(\,\sum_{abf} c_a\,{\bf
  w}\,\Bigr)\otimes\Bigl(\,\sum_{a^\prime b^\prime f^\prime}
c_{a^\prime}^\ast\,{\bf w}^\ast\,\Bigr)\,.\label{eq:wmatnew} 
\end{equation}

Note that it is absolutely essential to make the summation coherent
not only over the initial states (accounting for the non-diagonal
density matrix elements $c_a\,c_{a^\prime}^\ast$), but also use {\it
  coherent summation over the final states}. This rule is demanded
both by a theoretical selection rule (cf. Sect.~\ref{sec:clos}) and by
the experimental data. If it is obeyed, then 
we can reproduce in great detail the results of the laboratory
experiment, while if it is disobeyed we get zero polarization in the
corresponding test cases. The selection rule in
Sect.~\ref{sec:clos} implies that there is a strict coupling between the initial
and final state coherences. 

Since in this case each sum contains a much larger number of terms,
the physical system will contain a much richer set of coherences and
quantum interference phenomena, which are prevented in the earlier
approach to quantum scattering as represented by
Eq.~(\ref{eq:wmatold}). The new 
terms allow us to theoretically reproduce the results of our
laboratory experiment in great quantitative detail in situations when 
the old formalism predicts zero polarization. 

The practical implementation of the formally simple and symmetric
Eq.~(\ref{eq:wmatnew}) is however difficult, both conceptually and
technically, since we need a way to determine the phase relations
between the different ground level amplitudes $c_a$. In the absence of
a radiation field the relative phases are random. In the presence of
an incident radiation field the transverse oscillations of the
electric vector drive the dipole resonators of the atomic system. It
is this synchronized driving that produces phase locking between
combinations of $c_a$ level 
amplitudes. The heart of the problem is how to deal with this phase 
locking, which must be distinguished from conventional optical
pumping, although it may 
superficially resemble it. To conceptually understand the fundamental distinction
between phase locking and optical pumping we need a non-traditional
way to look at the nature of quantum transitions, an issue that will
be addressed in the next subsection. 

\begin{figure}
\resizebox{\hsize}{!}{\includegraphics{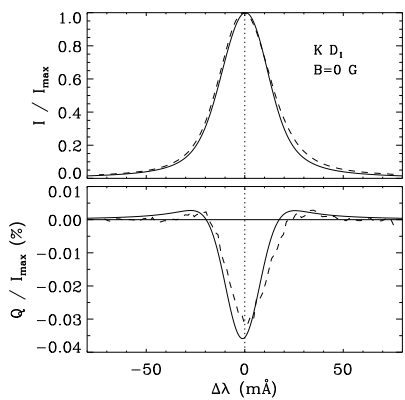}}
\caption{Polarized D$_1$ scattering by potassium when
  the incident radiation is 100\,\%\ linearly polarized perpendicular
  to the scattering plane ($Q/I=1$). The radiation scattered at $90^\circ$ is
  then partially polarized {\it parallel} to the scattering plane
  (output $Q$ negative). The observations (dashed curves) can
  be reproduced by the theory (solid curves) when
  the coherences in both the initial {\it and} final states are accounted for as in
  Eq.~(\ref{eq:wmatnew}). If the coherences in the final
  state would be 
  neglected, we would get zero polarization for all wavelengths, regardless
  of how unbalanced the $m$ state populations are. 
}\label{fig:d1qq}
\end{figure}

Before proceeding with the theoretical issues, let us already at
this stage present experimental evidence that backs up our previous
statements. In Fig.~\ref{fig:d1qq} we show the case of 90$^\circ$
scattering of K D$_1$ radiation when the incident light is
100\,\%\ linearly polarized perpendicular to the plane of scattering
(incident Stokes $Q/I=1$), and we measure Stokes $Q$ of the scattered
radiation. The laser has been tuned across the D$_1$ resonance, which
gives us 
the profile shape. Standard scattering theory
(Eq.~(\ref{eq:wmatold})) predicts exactly zero $Q$ polarization for
all wavelengths, while with Eq.~(\ref{eq:wmatnew}) we are able to
reproduce the observations in great quantitative detail. Note that the
observed $Q$ is negative, i.e., the scattered radiation is weakliy
linearly polarized {\it parallel} to the scattering plane. 

In the following sections we will explain what explicitly goes into
these calculations, how the theory is extended to include the new
physical effects, and how observations of the well-behaved D$_2$
system are used to constrain the theory and enforce
consistency. An objective is to provide convincing evidence that quantum
scattering has a far richer coherency structure than allowed by the old formalism of
Eq.~(\ref{eq:wmatold}), and to indicate the direction in which the
theory should be extended to cover these phenomena. 

To conclude this subsection we give the well-known expression for how to
obtain the Mueller matrix {\bf M} that describes scattering of the
Stokes vector from the complex coherency matrix {\bf W}: 
\begin{equation}
{\bf M}={\bf T}\,{\bf W}\,{\bf T}^{-1}\,.\label{eq:mueller}
\end{equation}
The complex $4\times 4$ matrices $\bf T$, ${\bf T}^{-1}$ in Eq.\~
(\ref{eq:mueller}) are purely mathematical transformation matrices
without physical contents \citep[cf.][]{stenflo-book94}.

\subsection{Wave packet interaction: Quantum blip or radiation bath\,?}\label{sec:blip}
An obstacle to a more complete understanding of the interaction of the
atomic system with the radiation field has been the often prevailing
misconception that atomic transitions are instantaneous events,
quantum `blips'. To make our conceptual discussion of the
  interaction process as transparent as possible we will refrain from
 applying the second quantization formalism with the quantization of the
electromagnetic field and instead use a semi-classical approach
in which photons are treated as exponentially damped wave packets. 

The misconception of instantaneous quantum jumps has been amplified by the
circumstance that a theory for polarized optical pumping is only
available under the assumption that the incident radiation is
broadband in frequency \citep[the flat-spectrum approximation,
cf.~p.~257 in][]{stenflo-lanlan04}. Although the flat-spectrum
  assumption only requires the radiation to be spectrally flat over
  ranges comparable to the Doppler width of spectral lines, let us for
conceptual clarity first consider the extreme case of complete
flatness.  With Heisenberg's uncertainty principle 
$\Delta E \Delta t\approx\hbar$ such a case implies that $\Delta t
=0$, i.e., the impinging wave packets have zero coherence length. Such wave packets do not
exist in nature. Furthermore, this limitation contradicts the principle
that each wave packet can be represented as a linear
superposition of its monochromatic Fourier components. 

While the flat-spectrum approximation does not imply this extreme
case of zero coherence length, it does imply wave packets with a
coherence length that is about two orders of magnitude smaller than
actual wave packets, since the Doppler width is typically two orders
of magnitude larger than the radiative damping width. For this reason
the implications of this approximation are qualitatively of the kind
indicated by our discussion of completely flat radiation. Thus, in terms of
hypothetical (non-existing) broad-band wave packets 
the pumping scenario is a sequence of almost discrete pumping events, in
which each quantum jump due to radiative excitation is
nearly instantaneous. During the relatively long waiting times between each discrete event the
atomic system is subject to the depolarization effects of collisions
and magnetic fields \citep[cf. the more extended discussion of this
topic in][]{stenflo-stenflospw6}. 

It is only meaningful to talk about quantum `blips' when a
wave packet is recorded in a detector (collapse of the wave
function). Before this happens the wave packets are not small
entities but huge, with coherence lengths of several meters (in the
absence of collisional effects). The radiative 
damping constant of potassium D$_1$ and D$_2$ is approximately 
$\gamma=3.8\times 10^7$\,s$^{-1}$ (determined by the inverse life time
of the excited state in the collisionless case), implying that the damped emission process
takes place over a time of order $1/\gamma$, which is $10^7$
  times longer than the time scale for the oscillation of the electric
  vector. During this long time light
travels approximately 8\,m, corresponding to $10^7$ wavelengths. 

With this enormous separation of time scales wave packets represent (from
the point of view of the atomic system) a radiation bath. During the
interaction with the atomic system the various dipole resonances in the atom are
driven $10^7$ times by the electric field of the single wave packet. This
gives plenty of time to establish synchronization between the phases
of the various dipole oscillators represented by the potential
radiative transitions between lower and upper magnetic substates. In 
the case of the D$_1$ atomic system there are 36 different allowed
combinations of dipole transitions between lower and upper $m$
states. Since all of them are subject to the driving force from the same
external oscillating electric field (provided by a single incident wave packet),
phase synchronizations will be established, depending on the polarization
of the incident wave packet. 

Both the lower and upper $J=1/2$ states of the D$_1$ atomic system is split
due to the nuclear spin into total angular momentum states 
$F=1$ (with three magnetic substates) and $F=2$ (with five magnetic
substates). There are thus 8 magnetic substates in each of the lower and upper
state. Various combinations of the 36 allowed dipole transitions have
common $m$ states (either lower or upper ones). For instance, 
a $\Delta m=0$ transition from upper state $F=1,\,m=0$ couples to both
$F=1,\,m=0$ and $F=2,\,m=0$ in the lower state. Since both the upwards
and downwards transitions are jointly driven by the same oscillating
electric field, phase synchronization between the two $F=1,\,m=0$ and
$F=2,\,m=0$ lower states will be produced, which then leads to new
polarization effects through Eq.~(\ref{eq:wmatnew}). 

Let us for completeness mention here that the fully quantized
  formulation of polarized scattering that was pioneered by
  \citet{stenflo-bommier97a,stenflo-bommier97b} does not make use of
the flat-spectrum approximation. Although the method is general with
the potential of being extended to the multi-level case, it has so far
only been implemented for the two-level case.

\subsection{Coherences of the atomic system}\label{sec:coher}
We have in the previous subsection seen that the wave packet interaction
is not associated with instantaneous quantum jumps, but that the
atomic system with its various dipole resonators (36 in the D$_1$
case) is immersed in a radiation bath (provided by each single
wave packet), which drives the atomic oscillations and couples combinations
of oscillators. 

This viewpoint is largely prohibited when the flat-spectrum approximation is
used, as indicated in the previous section. There are nevertheless good reasons why
this approximation has represented the established approach, because
by removing it we open 
a `Pandora's box' of coherency phenomena that we do not know how to
properly deal with. However, by keeping the lid shut on this Pandora's
box we prevent access to the world of coherency phenomena in which a
solution of the D$_1$ enigma is to be found. 

In the present paper we open the lid to examine the phenomena in this
new world and try to show how they are responsible for the observed
polarization features in the laboratory experiment. In view of the
complexity of the interconnected coherences our initial approach has
to be heuristic and phenomenological in nature rather than
representing an attempt at a formulation of a new general
theory. Still we go to great lengths to make the phenomenological
extension of standard scattering theory well defined, internally
consistent, based on the soundest possible arguments, and avoiding
free parameters as far as possible. This effort is only possible with
the guidance of experimental data. The success of our extension is
that it is capable of modeling the key experimental data in great
quantitative detail. 

In the present subsection we will examine the nature
of the various new coherences. For a conceptual understanding we begin
by examining how phase synchronization or phase locking plays out in
classical physics.

\subsubsection{Coherences in classical scattering}\label{sec:classic}
The equation for a classical oscillator in an external magnetic field
can be written as 
\begin{equation}
{{\rm d}^2 r_q\over{\rm d}t^2}-(2qi\omega_L-\gamma)
{{\rm d}r_q\over{\rm d}t}+\omega_0^2 r_q=-{e\over m}E_q
\label{eq:osccomp}
\end{equation}
in component form, when instead of the three Cartesian spatial
coordinates the three complex spherical vector components, labeled
with $q=0,\pm 1$, are used. The $q=0$ component is aligned along the
magnetic field. $r_q$ is the component of the position vector of the
oscillator, $\omega_L =eB/(2m)$ is the Larmor frequency. $\gamma$
represents the damping due to the radiative energy loss of the
oscillating system, and $\omega_0$ is the resonance frequency of the
system \citep[cf.][]{stenflo-book94}. 

The advantage of using the complex spherical vectors
rather than real linear vectors is that the component equations then
decouple and represent three independent oscillators. Still, although
the three equations are mathematically uncoupled (the equation system
is diagonalized), the three oscillators are phase locked because the
three components $E_q$ of the driving oscillating electric field of the
external radiation bath are mutually phase locked (since they are
components of the same oscillating vector). 

The solution of the stationary equation, when the oscillator is
immersed in an eternal, monochromatic radiation bath, leads to the
profile function $\Phi_{-q}$, which is proportional to the complex
refractive index of the medium. As the dipole radiation emitted by these
oscillators represents the scattered light, we readily obtain the
Jones scattering matrix in the form of Eq.~(\ref{eq:scatclass}). 

The frequency redistribution, i.e., the relation between the incident
and scattered frequencies, can be obtained by solving
Eq.~(\ref{eq:osccomp}) as a time-dependent problem
\citep{stenflo-bs99}. The full solution 
of this problem is the sum of the stationary solution and a transitory
solution. The collisional effects are generally described in two ways:
As phenomenologically introduced branching ratios between the
stationary and transitory solutions, and as phase truncations of the
oscillators, which lead to depolarization. Here we will not deal with
the details of such redistribution effects, since they are not crucial
for the key issues in the extension of scattering theory. 

However, one aspect that is fundamental to this extension and which
can be correctly described by the classical theory is the partial
decoherence of jointly driven oscillators that have different resonant
frequencies. In the absence of magnetic fields, the three classical
oscillator components have the same resonant frequencies and therefore
remain in phase for the duration of the radiation bath, there is no
decoherence. When however there is a frequency split induced by the Larmor frequency,
the oscillators get progressively out of phase, depending on the ratio
between the time scales of Larmor precession and radiative decay. This
partial decoherence leads to the polarization effects in the scattered
radiation that are covered by the term `Hanle effect'. 

The partial decoherence between jointly driven resonators, i.e.,
between jointly driven atomic dipole transitions, is by no means
limited to decoherence between the magnetically split transitions
(Hanle effect), but applies to the relative phase relations and
decoherence for any split but jointly driven transitions. An
example is the dramatic polarization effects seen on the Sun due to
scattering at the Ca\,{\sc ii} H and K line system \citep{stenflo-s80}, two lines of the
D$_1$ -- D$_2$ type (although without hyperfine structure). In this
particular case the line separation is due to the fine structure
splitting (different $J$ quantum numbers). In the potassium case that
we are dealing with the splitting is due to hyperfine structure
(different $F$ quantum numbers), and for non-zero magnetic fields due
to Zeeman splitting as well. 

When forming the bilinear products
$w_{\alpha\beta}\,w_{\alpha^\prime\beta^\prime}^\ast$ between the
elements of the Jones matrix for the scattering probability
amplitudes we get interference terms, in which the partial decoherence
is contained in the products $\Phi_{ba}\Phi_{b^\prime a^\prime}^\ast$
between the profile factors. An elegant mathematical transformation is
to convert such profile products into sums: 
\begin{equation}
\Phi_{ba}\Phi_{b^\prime a^\prime}^\ast\,\sim\,\cos\alpha
\,\,e^{i\alpha}\,\,{\textstyle {1\over 2}}\,(\Phi_{ba} +\Phi_{b^\prime
  a^\prime}^\ast)
\label{eq:phiqprod}
\end{equation}
\citep{stenflo-book94,stenflo-s98}, 
where the angle $\alpha$ is given by
\begin{equation}
\tan\alpha={\omega_{ba}-\omega_{b^\prime a^\prime}\over \gamma}\,.
\label{eq:tana}\end{equation}
In the case of magnetically induced splitting $\alpha$ is referred to
as the `Hanle angle'. The expression is however quite general and
is valid regardless of the physical origin of the splitting. 

With the conversion of Eq.~(\ref{eq:phiqprod}) the interference
effects due to the profile product gets factorized into a
frequency-independent Hanle part $\cos\alpha
\,\,e^{i\alpha}$ and a frequency-dependent part, which becomes unity
if we integrate over all frequencies (since each profile function is
area normalized). The Hanle part consists of an amplitude factor,
which in the scattering polarization 
corresponds to depolarization, and a phase factor that corresponds to rotation
of the plane of polarization. 

The classical oscillator case corresponds to the quantum-mechanical
case for a $J=0\to 1\to 0$ scattering transition. It also consists
of three resonators, which link the three $m$ state of the upper level
to the common, single lower state. In such a situation $a^\prime =a$,
so there can be no lower-level coherences. In the general case when
there are more than one lower state, ground-state coherences are
possible, but since the coherency terms not only contain the profile
products of Eq.~(\ref{eq:phiqprod}) but also the level amplitude
products $c_a c_{a^\prime}^\ast$, these coherences will only play a
role if there are correlations between the phases of certain
combinations of level amplitudes. In the absence of an external
radiation field the phases are uncorrelated, because there is no
coupling between the levels. Remaining lower-level coherences will be
erased if the time between collisions is much shorter than the
`waiting time' between succssive radiative excitations. The
needed synchronization between the 
phases can however be provided by the oscillating electric field of
the wave packet radiation bath, which is a source of coupling or phase locking
between combinations of ground states.

\subsubsection{Coherences in the initial and final states}\label{sec:cohinfin}
The coherences in the initial state are represented by the
off-diagonal elements $c_a c_{a^\prime}^\ast$  of the density
matrix. In the absence of outside disturbances (external radiation
field or collisions), each level amplitude oscillates like $\exp(-i
E_a\,t/\hbar)$. If the energy levels $E_a$ and $E_{a^\prime}$ are not
identical, the oscillations will rapidly get out of phase. Let us now 
assume that the phases get synchronized by the driving oscillating electric
field of an incident wave packet, but that the initial phase
synchronization decays with the damping rate $\gamma$ because the
driving electromagnetic wave packet is exponentially damped according
to this rate. In the time domain the density matrix element is then
$\exp(-i\, \omega_{aa^\prime} \,t - \gamma\, t)$. The time derivative of
this expression must for stationarity be balanced by the radiative
absorption and emission rates, which represent the interaction between the driving
oscillating electric field and the atomic system. The balance equation
then gives us the expression $1/(i\, \omega_{aa^\prime}  + \gamma )$ for
the corresponding density matrix element, which may more elegantly be
written as 
\begin{equation}
c_a c_{a^\prime}^\ast\,\sim\,{1\over 1+i\tan\beta}\,=\,\cos\beta\,\,e^{-i\beta}
\label{eq:caca}\end{equation}
\citep[cf. Eq.~(8.89) in][]{stenflo-book94}. The angle $\beta$ is given by
\begin{equation}
\tan\beta={\omega_{aa^\prime}\over \gamma}\,.
\label{eq:tanb}\end{equation}
Comparison with Eqs.~(\ref{eq:phiqprod}) and (\ref{eq:tana}) shows that
this initial-state coherence is of the same general form as 'Hanle
type' coherences between resonant transitions. 

If the oscillating electric field of the incident radiation has a
polarization that leads to radiative coupling between the $a$ and
$a^\prime$ states (via appropriate excited states), then we expect the
constant of proportionality in Eq.~(\ref{eq:caca}) to be close to
$\vert c_a\vert\,\,\vert c_{a^\prime}\vert$ (the product between
  the square roots of the sublevel populations), since when the phases
of the level amplitudes are synchronized in the radiation bath, the
only source of decoherence is the difference in oscillation frequency
as expressed by the Hanle-type factor in Eq.~(\ref{eq:caca}). In
the absence of any radiative coupling the proportionality constant
would be zero, with the consequence that the
respective initial-state coherence does not contribute to the
scattering process. In the next subsection (Sect.~\ref{sec:lockd1}) we
will discuss the nature of this radiative coupling in some more detail
using our laboratory experiment for potassium D$_1$ -- D$_2$ as an
example. 

As the energy separation between the $a$ and $a^\prime$ levels
increases, the initial-state coherence terms get suppressed towards
zero by a
`depolarization factor'  $\cos^2\beta$ represented by the real part of the
Eq.~(\ref{eq:caca}) expression. Because of
the symmetry between the radiative excitation and deexcitation
process, we expect there to be a similar suppression factor governing
the final-state coherences. There is however a difference between the
two legs of the $a\to b\to f$ scattering process: the $a\to b$ leg is
associated with the complex $\Phi_{ba}$ profile function of
Eq.~(\ref{eq:phimmu}) while the $b\to f$ leg is not. The imaginary
part of Eq.~(\ref{eq:caca}) with its associated minus sign is related
to the role of the complex profile function in the oscillation
  between the $a$ and $b$ states, which establishes the
radiative coupling. No such complex-valued term is involved in the
$b\to f$ leg (since the transition amplitudes $t_{fb}$ are real
numbers). Therefore it is most reasonable to expect that the
final-state coherences should be governed by the symmetric version of
Eq.~(\ref{eq:caca}), namely the sum of this expression with its
complex conjugate (normalized to unity for zero splitting). This
normalized sum equals the $\cos^2\beta$ suppression factor, with the
difference that now $\beta$ is not given by Eq.~(\ref{eq:tanb}), but
by the same equation if we replace $\omega_{aa^\prime}$ by
$\omega_{ff^\prime}$. 

The only valid benchmark that we can use to test the validity of these
rather heuristic arguments is experimental data. Fortunately the 
scattering polarization turns out to be extremely sensitive to the
choices we make for the ground-state coherences. If we omit them, we
get zero polarization, regardless of the relative initial $m$ state
populations. If we use Eq.~(\ref{eq:caca}) but set the final-state
coherences to be unity (no suppression factor), we get much too large
polarization effects. If we keep an imaginary part of the final-state
coherences like for the initial state, then we also cannot achieve a
fit with the data. The same happens if we omit the
imaginary term in Eq.~(\ref{eq:caca}) or change its sign. Also we must
only use the correct  
combinations of radiative couplings (cf. Sect.~\ref{sec:lockd1}), 
otherwise we get results that have no resemblance with the
observations. Only with the previously described choices of
expressions for the coherence terms we succeed in obtaining a fit at
all, and on top of that, the fit agrees 
with the observational data in great quantitative detail in a way that is
consistent between D$_1$ and D$_2$ and between the Stokes $I$, $Q$,
and $V$ parameters, including their profile shapes. 

Let us note here that for a given initial-state ($a,\,a^\prime$)
combination only certain ($f,\,f^\prime$) combinations for the
final-state coherences are allowed. The nature of these selection
rules will be addressed in Sect.~\ref{sec:clos}. Without them we find no
agreement with the observational data.

\subsubsection{Radiative coupling in the D$_1$ -- D$_2$ experiment}\label{sec:lockd1}
Our analysis here of the laboratory scattering experiment will focus
on the cases when (1) the incident radiation is 100\,\%\ linearly polarized
perpendicular to the scattering plane, while the Stokes $I$ and $Q$
parameters of the $90^\circ$ scattered
radiation are recorded, and (2) when the incident radiation is 100\,\%\
circularly polarized, while the Stokes $I$ and $V$
parameters of the scattered radiation are recorded. The
imposed external magnetic field is chosen to be vertical
(perpendicular to the scattering plane). This direction represents the
quantization direction in our theoretical discussion. 

Let us first turn our attention to case (1). Then among all the 36
resonators in the D$_1$ system only 
those with $m_a -m_b =q=0$ get driven by the oscillations of the
electric field of the incident wave packet, with the consequence that there
can only be radiative 
coupling between initial $a$ and $a^\prime$ states for which the $m$
quantum number is identical ($m_{a^\prime}=m_a$). The only possibility
for two such states to represent separate states is to belong to
different $F$ states. 

Let us as a concrete example see how the two states ($F=1,\,m=-1$) and
($F=2,\,m=-1$) are radiatively coupled. Each of them 
couple by radiative excitation to each of the two corresponding states
of the upper level. Similarly each of
the two upper states are coupled to the corresponding two lower states
by the emission process. The oscillating electric field drives the
upwards and downwards oscillating transitions with equal transition 
strength. 

All the four $m=-1$ states (two lower and two upper) are therefore
mutually phase locked as a consequence of the common driving external electric field. For
the same reason that we have phase synchronization between the
three classical oscillators (which leads to the Hanle factor of
Eq.~(\ref{eq:phiqprod}) for the upper states of a $J=0\to 1$
transition), the oscillatory symmetry between the upper and lower
states imply the presence of a corresponding phase synchronization between
pairs of ($a,\,a^\prime$) states with the Hanle type factor
of Eq.~(\ref{eq:caca}). 

For our case (1) we therefore set the
proportionality constant in Eq.~(\ref{eq:caca}) to its maximum
  value that represents full phase synchronization whenever
$m_{a^\prime}=m_a$, zero otherwise. This choice brings excellent
agreement between theory and observations, as illustrated in
Figs.~\ref{fig:d1qq} and \ref{fig:d1qbdep}. If we would set the
proportionality constant to its maximum value for all ($a,\,a^\prime$)
combinations we would get large positive $Q$ polarization, in
contradiction with the laboratory experiment, while omitting the
ground-state coherences leads to exactly zero for all wavelengths in
Fig.~\ref{fig:d1qq}. 

Now let us consider our case (2), when the incident radiation is
100\,\%\ circularly polarized and the
magnetic field is transverse. In this 
case the electric field of the incident wave packets oscillates with equal
amplitudes parallel and perpendicular to the magnetic field, which
means that both the $q=0$ and $q=\pm 1$ resonators of the atomic system
are excited. This leads to radiative coupling between {\it all} the
resonators and therefore between all the ($a,\,a^\prime$)
combinations. Consequently we should set the constant of
proportionality in Eq.~(\ref{eq:caca}) to its maximum value for all these 
combinations. Doing this we indeed get excellent agreement between
theory and experimental data, as illustrated in
Fig.~\ref{fig:d1vbdep}, while other choices do not lead to comparable
success. If we remove all ground-state coherences, we get an
order of magnitude too small polarization. This residual polarization is
exclusively due to quantum 
interferences in the excited state.

\subsection{Phase closure as selection rule}\label{sec:clos}
The ground-state coherences that govern the polarization of the
scattered radiation via Eq.~(\ref{eq:wmatnew}) have contributions from
both the initial and final states. In the previous section we have
indicated how the initial-state coherences are determined by the
radiative coupling between the initial $m$ states. In the present
subsection we will show that all final-state interferences are not
allowed but that they are governed by selection rules that couple them
to certain combinations of initial-state coherences. Therefore an 
initial-state coherence will only contribute to the
polarization of the scattered radiation if it is coupled in the
scattering process to a final-state coherence in a way that obeys the
selection rule. 

The selection rule that governs which combinations of initial and
final $m$ states are allowed can be formulated as a requirement of
{\it phase closure}, a concept that was introduced in
\citet{stenflo-stenflospw5}. It has its origin in the geometrical 
$\varepsilon$ factors in Eq.~(\ref{eq:scatgen}), which represent the scalar
product between the transverse radiation-field basis vectors ${\bf
  e}_{\alpha,\,\beta}$ and the 
complex spherical basis vectors ${\bf e}_q$ (with $q=0,\pm 1$) of the
atomic system. 

The basis vectors ${\bf e}_q$ are defined such that ${\bf e}_0$ is
along the magnetic field or quantization axis. In this coordinate
system the scattering geometry is specified by colatitude $\theta$ and
azimuth $\phi$, defined with respect to some arbitrarily chosen zero
point. The scattering geometry of our laboratory experiment is
  such that $\theta =90^\circ$ for both the incident and scattered
  beam, while the scattering angle is given by $\phi^\prime-\phi$, the
azimuth difference between the scattered and incident radiation. All
the azimuth  dependence is contained in $\varepsilon^\alpha_\pm$ and
$\varepsilon^\beta_\pm$ in the form of a phase factor $\exp(\pm
i\phi^\prime)$ or  $\exp(\pm i\phi)$,
regardless of the values of $\alpha$ and $\beta$ \citep[cf. Eq.~(3.86)
in][]{stenflo-book94}. The complex value of this phase 
factor depends on the arbitrary choice of zero point for the azimuths
in the coordinate system used. 

It would be unphysical if the polarization of the scattered radiation
would in any way depend on this choice of coordinate system. Therefore
only combinations of $\varepsilon$ factors for which the choice of zero
point cancels out can represent physical scattering processes. This
gives us a selection rule that we refer to as the requirement of phase
closure. The zero point of the azimuth scale cancels out if all the physical
effects depend exclusively on the {\it difference} $\phi^\prime-\phi$
between the scattered and incident azimuth directions, but never on
the absolute value of either azimuth. 

Each term in the Mueller scattering matrix is made up of the bilinear
products between scattering amplitudes because of the
tensor product in Eq.~(\ref{eq:wmatnew}). Since each scattering
amplitude contains two $\varepsilon$ factors, since 
$\varepsilon_q^{\beta\ast}=-\varepsilon_{-q}^\beta$ when $q=\pm 1$,
and since $q$
represents the difference between the lower and upper $m$ quantum
numbers of the transition, each Mueller matrix element contains four
$\varepsilon$ factors of the type 
\begin{equation}
\varepsilon^\beta_{m_a-m_b}\,\,\varepsilon^\alpha_{m_b-m_f}\,\,\varepsilon^\alpha_{m_{f^\prime}-m_{b\prime}}\,\,\varepsilon^\beta_{m_{b^\prime}-m_{a^\prime}}\,.
\label{eq:4eps}\end{equation}
$\varepsilon$ factors relating to the $a\to b$ transition contain the
azimuth $\phi$ for the incident radiation, while $\varepsilon$ factors
relating to the $b\to f$ transition contain the azimuth $\phi^\prime$
for the scattered radiation. If the combined complex phase factor for
our product of four $\varepsilon$ factors is $\exp(i\delta)$, then we
see from Eq.~(\ref{eq:4eps}) that 
\begin{eqnarray}
&\delta=(m_a-m_b)\,\phi +(m_b-m_f)\,\phi^\prime& \nonumber\\ &+(m_{f^\prime}-m_{b\prime})\,\phi^\prime +(m_{b^\prime}-m_{a^\prime})\,\phi\,.&
\label{eq:phasedelta}\end{eqnarray}

If we would change the zero point and replace $\phi$ and $\phi^\prime$
with $\phi-\phi_0$ and $\phi^\prime-\phi_0$, we see that the
dependence of the phase on $\phi_0$ only disappears if the factor in
front of $\phi$ equals the negative of the factor in front of
$\phi^\prime$. This gives us the selection rule 
\begin{equation}
m_f-m_{f^\prime}\,=\,m_a-m_{a^\prime}\,.
\label{eq:mselrule}\end{equation}
This phase closure relation thus demands that only the final-state
coherences that have the same $\Delta m$ as the initial-state
coherences are allowed. Note that the magnetic quantum numbers of
the excited state are not involved at all, the rule only dictates how
the initial and final state coherences are coupled. 

In our efforts to reproduce the observational data we
discovered an additional selection rule that seems to govern the
coupling between the initial and final state coherences: 
\begin{equation}
F_f-F_{f^\prime}\,=\,F_a-F_{a^\prime}\,.
\label{eq:fselrule}\end{equation}
It was not anticipated and only found by trial and error: With it we
got excellent fits to the observational data, without it the
disagreement with the data was very large. With afterthought
Eq.~(\ref{eq:fselrule}) is sensible, since then the scattering
transition radiatively couples only coherences that oscillate in the same
sense. A coherence between $F=1$ and 2, for instance, oscillates like
$\exp[-i(E_1-E_2)\,t/\hbar\,]$. A coherence with the reversed order of the
states would oscillate in anti-phase. Symmetry of the scattering
process seems to demand that couplings with anti-phase oscillations
are prohibited. At present we lack a full understanding of this rule,
but it is empirically required to get agreement with the observations.

\section{Theoretical modeling of the experimental data}\label{sec:model}
Although the calculation of scattering amplitudes from
Eq.~(\ref{eq:scatgen}) is well-known and uncontroversial, it is
technically difficult for our D$_1$ -- D$_2$ case, because we have
hyperfine structure splitting and are in the Paschen-Back regime, in
which the Zeeman splittings and the transition amplitudes vary with
field strength in a non-linear way. In support of the present project
S.V. Berdyugina and D.M. Fluri (priv. comm.) kindly provided
us with an IDL routine that they had written, which calculates the set of transition
amplitudes, resonance wavelengths, and energy levels for any given field
strength in the case of the potassium or sodium D$_1$ or D$_2$
system. The correctness of the program has been tested through
comparison with other codes as part of the work of
\citet{stenflo-sowmya2014a}. With this 
program the calculation of scattering amplitudes according to 
Eq.~(\ref{eq:scatgen}) becomes a straightforward task. 

When trying to theoretically reproduce the observational data from our
laboratory experiment there are three other effects to be
considered, which we have not discussed so far: (1) Effects of
collisions on the scattering polarization (will be dealt with in
Sect.~\ref{sec:coll}). (2) Optical depth effects with magnetically
induced dichroism (will be dealt with in Sect.~\ref{sec:tau}). (3)
Optical pumping of the lower $m$ state populations. 

While (1) and (2) will be accounted for, we choose to leave (3)
aside in our modeling presentation here. The reason is that optical
pumping that creates unbalanced $m$ state populations only modifies
the results quantitatively by modest amounts, not qualitatively, in
contrast to the fundamental issues that we have addressed in the previous
sections concerning the treatment of the initial and final state
coherences and the selection rules, all of which have dramatic effects on the
modeling. Since the `waiting time' between successive radiative
absorption events is in general much longer than the time between
collisions with the argon buffer gas, much of the population
imbalances can be expected to be erased before each scattering event. 

We nevertheless did modeling experiments also with inclusion of 
optical pumping that gives us 
the populations $\vert\, c_a\vert^2$, but even when the effects of
collisions were entirely disregarded in the pumping problem the
effects on the scattering polarization were not dramatic. It was 
easier to find a good fit to the observations when such pumping was
not included. Pumping does not play an
essential role in the context of identifying the missing physics in
polarized scattering. Including it introduces new free
parameters because the effect of collisions on the pumping process,
which certainly plays a significant role, is not sufficiently
known. For such reasons attempts to properly deal with optical pumping
would distract from the 
main issues of the paper and not lead to improved fits of the
data.

\subsection{Effects of collisions}\label{sec:coll}
The vapor cell used for our scattering experiment is small and
very convenient to handle. A heating coil in the stem of the cell
vaporizes the metallic potassium to produce the gas in the cell center
at which the scattering takes place. With a gas temperature
of 100$^\circ$ C we get high enough number density of potassium atoms
to provide the scattered light needed   
to measure small degrees of polarization. 

To suppress diffusion of the potassium gas, which could lead to
deposits on the cooler entrance and exit windows and make them opaque,
an argon buffer gas is used in the cell. Unfortunately this has the disadvantage that the
potassium atoms are subject to a high collision rate, which 
affects the scattering polarization in significant ways. 

The collisions cause 
line broadening by changing the effective damping constant of the
profile functions, as well as depolarization of the emitted radiation. Let
us, as is customary, denote the collisional damping 
constant by $\gamma_c$, while the natural, radiative damping constant
is $\gamma_N$. Then the effective $\gamma$ that governs the line shape
in  Eq.~(\ref{eq:phimmu}) is $\gamma_N + \gamma_c$. 

The collisional depolarization effects, which destroy the phase
relations in the scattering process to make the scattered light unpolarized and
isotropic can be expressed in terms of the depolarization factor 
\begin{equation}
k_c ={\gamma_N\over \gamma_N +D^{(K)}}\,,
\label{eq:kc}\end{equation}
where $D^{(K)}$ with $K=1$ or 2 represents the destruction rates of the
$2K$ multipoles of the atomic system. In time-dependent classical
scattering theory \citep{stenflo-bs99} 
\begin{equation}
D^{(K)}=0.5\,\gamma_c
\label{eq:dk}\end{equation}
for both $K=1$ (atomic orientation) and $K=2$ (atomic alignment). In
the classical theory the collisions truncate the exponentially damped
atomic oscillations. After Fourier transformation we then get the factor
of 0.5 in Eq.~(\ref{eq:dk}). In detailed quantum mechanical
calculations of the scattering process the resulting value for this
factor depends on the atomic system but stays typically within at
least a factor of two around the classical value. In the absence of
other data our best choice is therefore to use the relation
represented by Eq.~(\ref{eq:dk}). 

The depolarization factor $k_c$ is used to scale all elements of the
Mueller scattering matrix except the first one ($M_{11}$), which represents isotropic
scattering. This is a simplified
treatment, but in view of the large collisional rates and the use of
these rates for a parametrization of the problem, it is not justified
here to go into the intricacies of partial frequency redistribution
theory. 

The suppression effect from the depolarization caused by $k_c$ is largely compensated
for by the increased overlap and interference between the ground-state
sublevels due to the collisional broadening, as described by
Eq.~(\ref{eq:tanb}). The question is however what collision-modified
expression for $\gamma$ we should use in Eq.~(\ref{eq:tanb}), e.g. $\gamma_N+\gamma_c$ or
$\gamma_N+D^{(K)}$\,?  On the one hand one may argue for the first
of these two choices, since these coherences are determined by the radiative
couplings between the various combinations of allowed transitions, and
these couplings are governed by the profile functions in which
$\gamma=\gamma_N+\gamma_c$. On the other hand there is a good case for
the second expression, since the depolarization factor used for the
upper-state coherences is based on $D^{(K)}$. Through experimentation
with both versions we find that the best fit results are 
obtained if we use an average between the two cases. This implies that
we for the effective $\gamma$ in Eq.~(\ref{eq:tanb}) use
$\gamma_N+0.75\gamma_c$. For symmetry reasons we use the same
expression for the initial and final state coherences. 

Let us stress that the choice of the expression for $\gamma$
to be used for the ground-state coherences is not crucial for our
investigation. It does not lead to any qualitative changes, but only affects
the quantitative details of the model fitting. Our aim here is to make
the choices that appear most reasonable while minimizing the number of
free parameters. 

If we thus fix the choices for the three relations used to describe
the effects of collisions, $\gamma_N+\gamma_c$ to describe the shape
of the profile function, $\gamma_N+0.5\gamma_c$ to describe the
collisional depolarization factor used for the Mueller matrix
elements, and $\gamma_N+0.75\gamma_c$ to describe the effect on the
ground-state coherences, then we are left with a single free
  collisional parameter, namely $\gamma_c$.

\subsection{Optical depth effects}\label{sec:tau}
The optical depth of the potassium gas must be non-zero, otherwise we
would not get any scattered photons. Choosing the working temperature
of the cell represents a compromise: we want to optimize the number of
scattered photons to get a good signal, while avoiding multiple
scattering or excessive optical depth effects. The finite
optical depth leads to enhanced absorption at line center, which may 
change the profile shape, and also causes 
dichroism when a magnetic field is present. Comparison between our
theory and the experimental data shows that the optical depth effects
are too small for significant multiple scattering, but large enough to
cause significant line broadening and magnetically induced
dichroism. They therefore need to be accounted for, but a single
  free parameter, the optical depth $\tau$ at line center (where the
optical depth reaches its maximum value) is sufficient. We find that the optimum fit
value for $\tau$ is 0.25 in the case of D$_1$, while it is twice this
value for D$_2$ because of its twice higher oscillator strength. 

Our scattering experiment can be described as a transformation of
the input Stokes vector ${\bf S} _{\rm in}$ into the output Stokes
vector ${\bf S} _{\rm out}$ through 
\begin{equation}
{\bf S _{\rm out}}={\bf M}_{\rm eff}\,\,{\bf S _{\rm in}}\,,
\label{eq:sinout}\end{equation}
where ${\bf M}_{\rm eff}$ is the effective Mueller matrix of the
system. When optical depth effects are accounted for, it can be
represented by 
\begin{equation}
{\bf M}_{\rm eff}={\bf M}_{\rm arm}\,\,{\bf M}_{\rm scat}\,\,{\bf M}_{\rm arm}\,,
\label{eq:muellereff}\end{equation}
where ${\bf M}_{\rm scat}$ is the scattering Mueller matrix, the same
as the ${\bf M}$ that is given by Eq.~(\ref{eq:mueller}), while ${\bf
  M}_{\rm arm}$ represents the Mueller matrix for the medium in an arm
of the cell. There are two relevant arms here: The incident light goes
through the input cell arm before it reaches the cell center, where it
is scattered and then traverses the output cell arm before being
analyzed for its polarization state and being detected. The Mueller
matrices for the input and output arms are the same and surround the
scattering matrix. 

In the unpolarized case with an optical depth $\tau$ at line center
and an absorption profile $\phi$ that is normalized to unity at line
center, the effect of the medium would be absorption by the factor
$\exp(-\phi\tau)$ (since the contributions from emission 
through scattering in the pencil-shaped geometry of the medium would
give a negligeable contribution 
in the particular beam direction in comparison with the absorption effects in
that direction). In the polarized case, when the medium in a cell arm
is described by the $4\times 4$ Mueller matrix ${\bf \Phi}$ instead of
by a scalar absorption coefficient, the dichroic effects of the medium are
represented by 
\begin{equation}
{\bf M}_{\rm arm}=e^{-{\bf \Phi}\,\tau}\,.
\label{eq:muellerarm}\end{equation}
The exponentiation of a matrix is defined by its Taylor expansion. 

In a magnetized medium 
\begin{eqnarray}
{\bf \Phi}=\!\left(\matrix{\phi_I &\phantom{-}\phi_Q &\phantom{-}\phi_U &
\phantom{-}\phi_V\nonumber\\ \phi_Q &
\phantom{-}\phi_I &\phantom{-}\psi_V &-\psi_U\nonumber\\ \phi_U &-\psi_V &
\phantom{-}\phi_I &\phantom{-}\psi_Q\nonumber\\ 
\phi_V &\phantom{-}\psi_U &-\psi_Q
&\phantom{-}\phi_I}\!\!\!\!\!\!\!\!\!\!\!\!\!\!\!\!\!\! \right)\,, 
\label{eq:phimat}\end{eqnarray}
where $\phi_{Q,U}$ describe the transverse Zeeman effect, $\phi_V$ the
longitudinal Zeeman effect, and $\psi_{Q,U,V}$ the magneto-optical
effects, while $\phi_I$ represents the effect of the Zeeman splitting
on the intensity absorption profile \citep[cf. Eqs.~(6.58) and (6.59)
in][]{stenflo-book94}.  

Since we are dealing with a very weakly magnetized medium (all field
strengths that we consider here are below 20\,G), all off-diagonal
elements of ${\bf \Phi}$ are small in comparison with $\phi_I$. In our
power series expansion of $\exp(-{\bf \Phi}\,\tau)$ we therefore only
need to retain the terms of first order in these elements, while
allowing the optical depth $\tau$ to be of any higher order. 

Here we will only model the case when the incident
radiation is propagating perpendicular to the magnetic field
direction, which also defines the positive Stokes $Q$ direction. Then
all dichroic elements of ${\bf \Phi}$ are zero except $\phi_I$ and 
$\phi_Q$. In the case of circularly polarized light $\phi_Q$ plays no
role, the whole effect is in the form $\exp(-\phi_I\,\tau)$. However, in
the case of linear $Q$ polarization, $\phi_Q$ is 
significant and represents the effects of magnetically induced
dichroism. 

Without this dichroism we are unable to explain the observed
dependence of the scattered $Q$ polarization on field strength, since
our scattering theory alone does not predict any significant effect,
even when the ground-state coherences are included. Comparison with
our data gives us an
estimate of the value of $\tau$ that is needed to explain the
observed field dependence. Another constraint comes from the observed
profile shapes. The line width increases with increasing $\tau$. It
turns out that the value of $\tau$ that gives us the right field
dependence automatically also gives us the right line width. 

For the calculation of the ${\bf \Phi}$ matrix we have to sum over the
contributions from the various $\phi_q$ profile functions ($q=0,\pm
1$) weighted by the corresponding transition strengths $t_{ab}^2$, to arrive
at the expressions for $\phi_I$ and $\phi_Q$. In this case the profile
functions $\phi_q$ are described by Voigt functions, having a Doppler width
given by the thermal width at 100$^\circ$ C (10.2\,m\AA) and a
damping constant given by $\gamma_N+\gamma_c$, which is the same
damping rate that governs the profile functions of the scattering
process.

\subsection{Using D$_2$ to determine the free parameters}\label{sec:d2par}
With our previous definitions we are left with only two free model parameters, $\gamma_c$,
which governs all the collisional effects, and $\tau$, which governs
the dichroic or optical-depth effects. For physical consistency we
further have the constraint that the same parameter values must be used
for {\it both} the D$_1$ and D$_2$ observations. We are
not allowed to use independent fit parameters for these two lines, 
since the same vapor cell is used for both. The only modification is
that we need to use $2\tau$ for D$_2$, since it has twice as large
oscillator strength and therefore twice as large optical thickness. 

The ground-state
coherence effects that we have been discussing play no significant
role for the D$_2$ line, because D$_2$ has intrinsic polarization (from
the upper level) that is larger by approximately a factor of 50 and
therefore masks the ground-state effects. For this reason the  D$_2$ line can perfectly well
be described by standard scattering theory. By using 
the D$_2$ line to fix the values of our two free parameters,
the obtained values become independent of the way in which
we are dealing with the ground-state coherences. No further parameter
fitting is then needed when addressing the D$_1$
problem. 

\begin{figure}
\resizebox{\hsize}{!}{\includegraphics{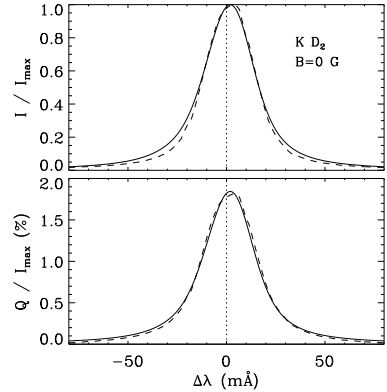}}
\caption{Stokes $I$ and $Q$ profiles for potassium D$_2$ in the
  non-magnetic case when the incident radiation is
  100\,\%\ linearly polarized in the $Q$ direction (perpendicular to
  the scattering plane) like in Fig.~\ref{fig:d1qq} that represented the
  D$_1$ case. The same values for the free parameters,
  $\gamma_c=90\,\gamma_N$ and $\tau=0.25$, are used for both the D$_2$
  and D$_1$ cases. Note that the D$_2$ polarization is about 50 times
  larger than the D$_1$ polarization and of opposite sign. 
}\label{fig:d2qq}
\end{figure}

Figure \ref{fig:d2qq} shows the excellent fit we get to the observed Stokes
$I$ and $Q$ D$_2$ profiles for the non-magnetic case when the incident
radiation is 100\,\%\ linearly polarized perpendicular to the
scattering plane ($Q$ type polarization). The dashed curves represent
the observations, the solid curves the theoretical model with
$\gamma_c =90\,\gamma_N\approx 3.5\times 10^9$\,s$^{-1}$ and $\tau=0.25$
(implying that the maximum 
optical depth at the D$_2$ line center is 0.50). The theoretical
profiles have been convolved with a Gaussian having a Doppler width
that equals the thermal width for 100$^\circ$ C. No additional
instrumental broadening has been applied (we have no reason to expect
any significant spectral smearing of the laser beam). 

The fit in Fig.~\ref{fig:d2qq} is nearly perfect,
except that there are some deviations in the distant wings of the $I$
profile. The $Q$ polarization amplitude scales directly 
with the depolarization factor $k_c$, which therefore gets fixed
precisely by the data. Through
Eqs.~(\ref{eq:kc}) and (\ref{eq:dk}) this determines the value of
$\gamma_c$. 

The line width scales with $\gamma_c$, but it also increases with
increasing $\tau$, since the absorption mainly suppresses the line core
while having little effect on the wings. If we use the $\gamma_c$ that
is fixed by the $Q$ amplitude but set $\tau=0$, then the theoretical
profiles are significantly narrower than the observed ones, and there
would be a need for additional line broadening (e.g. due to 
some hypothetical instrumental smearing) to get agreement
between the profile shapes. With our chosen value for $\tau$ we get an
excellent profile fit without the need for any such additional ad hoc 
broadening. 

As we will see in the next subsection, the $\tau$ that is needed to fit
the D$_2$ profile shape is also the $\tau$ that is needed to account
for the field strength dependence of the D$_1$ $Q$ polarization. It is
gratifying to note that there is such consistency between the
independent D$_2$ and D$_1$ constraints on $\tau$. 

If we would increase the value of $\tau$
significantly further, so that $2\tau$ would start to become
comparable to unity, then the D$_2$ Stokes $I$ profile not only
broadens but gets deformed by becoming double-humped with a depression
at line center. The absence of any sign of such a 
deformation shows that we can safely rule out the possibility that
the optical depth could be much larger than our present
estimates. This conclusion refers only to the
qualitative profile shape without using numerical values of any
amplitudes or widths.

\subsection{Role of magnetically induced dichroism}\label{sec:dichr}
In Fig.~\ref{fig:d1qq} we showed the results for D$_1$ scattering in
the non-magnetic case with the same geometry and same parameter values
as for the D$_2$ plot in Fig.~\ref{fig:d2qq}. We note that the D$_1$
polarization is of opposite sign and with about 50 times smaller
amplitude than that of $D_2$. The laboratory experiment has the
sensitivity to measure polarization effects down in the $10^{-5}$
range. Such a sensitivity is needed to uncover the tiny polarization
effects and profile shapes for D$_1$ scattering. The access to this
parameter range allows us to explore scattering physics in previously
untested domains. 

\begin{figure}
\resizebox{\hsize}{!}{\includegraphics{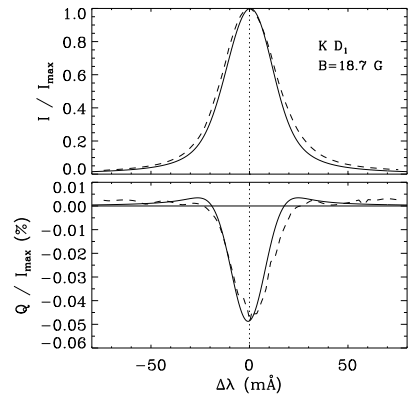}}
\caption{Stokes $I$ and $Q$ profiles for K D$_1$ for the same geometry
  and parameters as in Fig.~\ref{fig:d1qq}, except that here we show
  the case when the vertical magnetic field is 18.7\,G (while it was
  zero in Fig.~\ref{fig:d1qq}). Dashed curves: observations; solid
  curves: theory. The difference in polarization scale
  with respect to Fig.~\ref{fig:d1qq} is due to magnetically induced
  dichroism in the potassium gas. The systematic wavelength shift of
  the dashed $Q$ curve relative to the solid one can be understood in
  terms of the time constant used for the lock-in amplifier to
  optimize the polarimetric sensitivity. 
}\label{fig:d1qbdep}
\end{figure}

When for D$_1$ doing the Stokes $Q$ scattering for a sequence of different field
strengths (with the field oriented perpendicular to the scattering
plane) the laboratory experiment reveals a significant field
dependence that scales with the square of the field strength. In contrast, a
significant field dependence (within the studied field strength range)
is absent in the computed $Q$ polarization when the effects of the
surrounding medium are not accounted for. If we however introduce a
finite optical depth, then we
get a field dependence that also scales with the square of
the field strength, as it should when it is due to the transverse
Zeeman effect. The amplitude of this dependence
scales with $\tau$. Choosing $\tau=0.25$ gives good agreement
with the observational data, as illustrated by Fig.~\ref{fig:d1qbdep}. 

The main source of this field dependence comes from the
$\phi_Q$ element of the dichroic ${\bf \Phi}$ matrix of
Eq.~(\ref{eq:phimat}), which represents the transverse Zeeman effect. In absorption the
contribution from the $\pi$ component in the line core is negative, while the
$\sigma$ contributions in the wings are positive, but in the $\phi_Q$
profile they are
balanced to make the integrated profile zero. However, the dichroic
contribution to the $Q$ polarization comes not from 
$\phi_Q$ alone, but from its product with the intensity profile. This
gives much more weight to 
the negative contribution from the line core than to the wing
contributions. Therefore the net contribution is negative, with the
consequence that the already negative non-magnetic scattering $Q$ 
polarization gets enhanced. 

Figure \ref{fig:d1qbdep} shows the $I$ and $Q$ profiles for the
largest field strength used in our laboratory experiment, 18.7\,G. As
in Fig.~\ref{fig:d1qq} the observations are represented by the dashed
curves, the theory by the solid curves. The geometry and other
parameters are the same as for Fig.~\ref{fig:d1qq}, the only
difference being the field strength. We see that an excellent fit is
obtained, which would not have been the case without introducing the
finite optical depth $\tau=0.25$. Comparison between
Figs.~\ref{fig:d1qbdep} and \ref{fig:d1qq} shows that the profile
shapes are almost unaffected by the magnetic field, but the
polarization scales in the two figures are significantly different
(although the difference is rather modest). 

Both Figs.~\ref{fig:d1qbdep} and
\ref{fig:d1qq} give the (false) impression that the theoretical model
is unable to reproduce the exact wavelength position of the $Q$
profiles. The observational (dashed) curves for the $Q$ profiles are
seen to be systematically
shifted towards the right relative to the theoretical (solid)
curves. The magnitude of the shift is about 2.8\,m\AA, which
corresponds to about 1.7 step sizes in the stepwise wavelength tuning
of the laser to scan the line profile. 

The probable origin of this shift has to do with the time constant
used by the lock-in amplifier when demodulating the polarization
information, which for Stokes $Q$ is at 84\,kHz due to the
piezo-elastic modulation. To achieve polarimetric sensitivity down into
the $10^{-5}$ range one needs maximum temporal integration of the
demodulated AC signal before it is read out. The chosen analog
integration time should however be compatible with the scan rate. If
the integration time is longer than the time between tuning steps,
there will be a delay and overlap between the steps, resulting
in shift and broadening of the demodulated profile. As such an analog
time constant is not applied to the DC (Stokes $I$) signal, the $Q$
profile will be shifted (and slightly broadened) in the tuning
direction with respect to the $I$ profile. The observed shift that is
of order 1-2 tuning steps is a natural consequence of trying to
maximize the polarimetric sensitivity of the measurements. 

Note that the time constant could be much reduced in the D$_2$
case, since the polarization signals were about 50 times
larger. Therefore we do not see any corresponding shift of the $Q$
profile in the plot for D$_2$ in Fig.~\ref{fig:d2qq}.

\subsection{Circularly polarized scattering for transverse magnetic
  fields}\label{sec:circ}
Let us now turn to the D$_1$ case when 100\,\%\ right-handed circularly
polarized incident radiation is used, and Stokes $I$ and $V$ are
measured in the scattered radiation. The geometry remains the same as
in our previous discussion of $Q$ scattering, 
with the magnetic field being perpendicular to the scattering plane. 

\begin{figure}
\resizebox{\hsize}{!}{\includegraphics{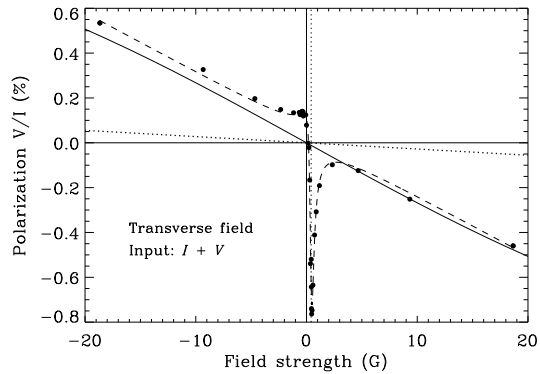}}
\caption{Degree of circular polarization in the beam scattered at
  $90^\circ$ by potassium  D$_1$, plotted 
  as a function of field strength. The fractional circular polarization
  here represents the area of the Stokes $V$ profile divided by the area of
  the $I$ profile. The incident beam has 100\,\%\ right-handed
  circular polarization, and the magnetic field is perpendicular to the
  scattering plane. The observations are represented by the filled
  circles and the dashed curve, the theoretical model that includes initial and
  final state coherences by the solid line. When these ground-state
  coherences are omitted but the interference terms of the
  excited state are retained, one obtains the dotted line, which has a slope that is
  an order of magnitude smaller. If all $m$ state coherences would be
  omitted the polarization would be exactly zero. 
}\label{fig:d1vbdep}
\end{figure}

The laboratory measurements were made for a sequence of field
strengths between $\pm 18.7$\,G. In Fig.~\ref{fig:d1vbdep} we
represent the results as the fractional circular polarization
$V/I$\,(\%) vs. the strength of the vertically oriented field. In this
case $V$ represents the wavelength integrated $V$, $I$ the integrated
Stokes $I$ profile. The 
observations are given by the filled circles and the dashed 
analytical fit function, while the solid line is obtained with our
theory, using the same values for parameters $\gamma_c$ and $\tau$ as
we used for the modeling of $Q$ scattering. In contrast to the linearly
polarized case, the incident circular polarization leads to radiative
couplings between all the initial $m$ states and not only between those with
identical $m$ quantum numbers, as explained in
Sect.~\ref{sec:lockd1}. 

We notice in Fig.~\ref{fig:d1vbdep} the excellent agreement between the
solid and dashed curves, in particular the equal slopes of the field
dependence, except for fields in the inner range 0-2\,G. The remarkable
negative peak is centered at the value 0.43\,G (marked by the vertical
dotted line in the diagram). This offset from zero
can be understood as being caused by the Earth's magnetic field,
which has a downwards pointing component of approximately this magnitude. To
compensate this component our imposed magnetic field needs to have an
upwards (positive component) of this magnitude, which leads to the
apparent small shift of our field strength scale. 

Since the Earth's magnetic field is not exactly vertical but inclined,
the residual field will become horizontal when the vertical field gets
compensated, but the azimuth orientation of our experimental setup with respect
to the azimuth of this residual horizontal component is not
sufficiently known to allow meaningful modeling of this behavior. In the
milligauss range in the immediate neighborhood of the $+0.43$\,G
offset we can expect the magnetic-field vector to rapidly change its
orientation between vertical and horizontal. The sharp negative peak
in Fig.~\ref{fig:d1vbdep} shows that the scattering polarization has
enormous sensitivity to the field orientation in this milligauss
range. The existence
of this sharp peak indicates that our description of the
physical system is still incomplete. The radiation-matter
interaction appears to have a much richer substructure than anticipated. 

Other laboratory
tests that we have made show that when we have horizontal magnetic
fields aligned along the propagation direction of the incident
radiation and choose this radiation to be 100\,\%\ circularly
polarized, then not only the scattered Stokes $V$, but also the
scattered Stokes $I$, varies dramatically with the strength of the
imposed field. In contrast to the vertical field case, for which 
Stokes $I$ is insensitive to field strength, the horizontal field case
leads to a 
Stokes $I$ that has a sharp peak around zero field strength and rapidly
declines towards zero as the field strength increases
\citep{stenflo-thalmannspw5}. This behavior can be 
readily understood in terms of extreme optical pumping of the
medium. With the described geometry either only $q=+1$ or $q=-1$
transitions will be excited (depending on the handedness of the
incident circular polarization). If for example only radiative
excitations that raise the $m$ quantum number by +1 are possible, then
all the atoms of the medium will quickly be pumped into the ground
state with $m=+2,\,F=2$, from which no more radiative excitation
is possible. This means that the medium becomes transparent and the
scattered intensity goes to zero, as observed except for the weak fields
around zero. The Stokes $I$ peak at zero
field strength occurs because there are weak stray fields 
(in particular the Earth's magnetic field), which break the symmetry and
prevent the extreme pumping scenario from happening. 

The laboratory experiment was also done with the opposite handedness
of the input polarization (100\,\%\ left-handed circularly
polarized radiation). The result is the same, except that the sign of
the output $V/I$ is reversed (and consequently the sharp peak at +0.43\,G is
positive). The exact sign reversal also happens for the theoretical
model, the diagram looks the same if we reflect it in the horizontal
axis. This symmetry behavior implies (as verified by our theory) that
all the $V$ signal in the output radiation comes from the non-zero
Mueller matrix element $M_{44}$, while the other elements of the 4th
row, like $M_{41}$, are exactly zero. 

In contrast to the scattering of $Q$ polarization, dichroism does not
play any role in Fig.~\ref{fig:d1vbdep}. The reason is that
the term $\phi_V$, which represents the longitudinal Zeeman effect in the
Mueller matrix ${\bf \Phi}$, is zero because we have
chosen the magnetic field to be transverse. Therefore the effect of the finite
optical depth is to multiply both the scattered $I$ and $V$ profiles
with the same factor, $\exp(-\phi_I\,\tau)$, which differentially
suppresses the line core more than the wings. Since it turns out that
the scattered $I$ and $V$ profiles have
identical shapes and thus are deformed in the same way by the factor
$\exp(-\phi_I\,\tau)$, the fractional polarization, integrated $V$
divided by integrated $I$, is
not affected. This is not so in the $Q$ scattering case, since the $I$
and $Q$ profiles have different shapes, and in addition there is the
contribution from $\phi_Q$, which is non-zero since it represents the
transverse Zeeman effect. 

If we would remove the initial and final state coherences but keep the
upper-state coherences (like in the standard scattering theory of
Eq.~(\ref{eq:wmatold})), then we obtain the dotted curve in
Fig.~\ref{fig:d1vbdep}, which has a slope that is an order of
magnitude smaller than that of the observations. If we would also
remove the upper-state interferences we
would get exactly zero for all field strengths. This demonstrates that
all the polarization effects in $V$ scattering for transverse
magnetic fields have their origin in coherences between separate
magnetic substates, and that the dominating contributions
come from coherences in the ground state.


\section{Limitations, conclusions, and outlook}\label{sec:conclude}
Our search for missing physics in quantum scattering was motivated by
the observed existence of a small but significant polarization peak in
the Sun's spectrum of the Na\,{\sc i} D$_1$ line at 5896\,\AA. Quantum
theory appeared to predict that the D$_1$ line should be intrinsically
unpolarizable, and the transverse Zeeman effect could be ruled out as a
source of the observed polarization. After a decade of failed attempts
to explain this enigmatic feature we decided to set up a laboratory
experiment to address the question whether this was a problem of solar
physics or quantum physics. 

Quantum systems often contain an inherently rich and complex structure that is
not always fully understood. The correct application of a complex 
theory needs guidance from experimental data. The problem is that
the theory of polarized scattering at multi-level atomic systems has
never been experimentally tested before in the parameter range that we
are addressing here. This needs to be done before we can have
sufficient 
confidence in the application of the theory to `messy' astrophysical
environments over which we have no control. 

In most cases when new types of instrumentation allows the exploration
of parameter domains that have never been accessed before, surprising
new phenomena have been discovered which were not predicted by the
theory available at the time. The D$_1$ enigma was never expected but
was a major surprise when the Sun's spectrum could be recorded with
high-precision imaging spectro-polarimetry allowing a sensitivity of
$10^{-5}$ in the fractional polarization. Laboratory experiments in
polarized scattering was a hot topic in the early days of quantum
mechanics but they were largely abandoned around 1935 in favor of other
topics. The polarization phenomena that we are discussing here are way
out of reach for the experimental technology available at that time. 

The aim of our laboratory experiment is not to try to emulate
scattering in the solar atmosphere, but to test the basics of D$_1$
scattering physics under simple and controlled conditions. The
experiment revealed that there is indeed a problem of quantum physics:
D$_1$ exhibits a rich 
polarization structure in situations where available theory predicts
zero polarization. 

When abandoning the prevailing `flat-spectrum approximation' we open a
Pandora's box to a new world of coherency phenomena that were
excluded before. In the present paper we show that polarized
scattering theory for multi-level atomic systems has to be extended to
include {\it both} the initial-state {\it and} final-state coherences
(in addition to the coherences in the excited, intermediate
state) in order to explain the observed polarization structures. With
our phenomenological 
extension of the theory we succeed in 
reproducing the observations not only qualitatively in terms of
profile shapes and signs, but also quantitatively in great detail,
both for the scattering of linear and circular
polarization in a transverse field geometry. This gives us confidence
that the missing ingredients of the theory have been identified
correctly. 

There are however also indications that our theoretical extension is
far from complete. Experiments with the scattering of circular
polarization indicate 
that there are dramatic effects happening in the milligauss regime of
the external magnetic field (cf. Fig.~\ref{fig:d1vbdep}). We are
presently unable to deal with this regime, because we lack control and
knowledge of the strength and orientation of the field in the
milligauss range due to the presence of stray fields, including the
Earth's magnetic field. An experiment that could properly deal with this
issue is likely to reveal additional substructures of the atomic
system that have not been identified so far. To fully address the
behavior for such weak fields the experiment needs to be set up in a
magnetically clean environment, which is not easy but doable. 

Another deficiency is that we only had access to a vapor cell that
used argon buffer gas to prevent condensations on the cooler cell
windows. As a consequence the potassium atoms experience a high
collision rate (with $\gamma_c\approx 
90\,\gamma_N$), which produce large line broadening and depolarization
effects. These effects had to be accounted for in a phenomenological and
parametrized way. In future one would like to do the experiment with a
cell that does not use any buffer gas, so that one can explore the clean case of scattering
in the collisionless domain. In the solar chromosphere, where the
core of the Na\,{\sc i} D$_1$ line is formed, one should approach
  nearly collisionless conditions, but according to 
  \citet{stenflo-kerkenibommier02} lower level polarization in
  Na\,{\sc i} D$_1$ will always be destroyed under solar conditions. Ideally
one would like an experimental 
setup that would allow exploration of the scattering polarization as a function of
collision rate, by being able to change and control the amount of
buffer gas being used. 

It is generally problematic to apply an established physical theory
to parameter domains in which the theory has never been tested by
experiment. We do not have a theory that has been sufficiently tested
experimentally when it comes to the interpretation of scattering
polarization from multi-level atomic systems. Experimental exploration of this
poorly studied domain may bring unexpected new insights into neglected
aspects of quantum physics. In the present paper we have tried to
indicate the direction in which the exploration of this domain could
take us.


\acknowledgments
The potassium cell was specially tailor made free of charge for our
laboratory experiment by the late Alessandro Cacciani, who had devoted
much of his life to perfect the art of making such cells for
magneto-optical filter systems used in helioseismology and solar magnetometry. The
scattering experiment was set up and carried out at ETH Zurich as part of two PhD
projects, by Alex Feller and Christian Thalmann. The data bank set up
by Thalmann for the collection of his laboratory recordings has served
as a resource to guide the theoretical efforts. I am grateful
to Svetlana Berdyugina and Dominique Fluri for providing me with their
IDL code to calculate the atomic level structure and transition
amplitudes in the Paschen-Back regime for the sodium and potassium
D$_1$ -- D$_2$ systems, and to Roberto Casini for various discussions
on quantum physics, in particular on the nature of the flat-spectrum approximation. 


\end{document}